# Dynamic Full-Field Optical Coherence Tomography module adapted to commercial microscopes for longitudinal *in vitro* cell culture study

Tual Monfort[1, 2, 3], Salvatore Azzollini[1], Jeremy Brogard[1], Marilou Clémençon[1], Amélie

Slembrouck-Brec[1], Valerie Forster[1], Serge Picaud[1], Olivier Goureau[1], Sacha Reichman[1],

Olivier Thouvenin[4*] and Kate Grieve[1, 2, 3*]

[1] Sorbonne Université, INSERM, CNRS, Institut de la Vision, 17 rue Moreau, F-75012 Paris, France
[2] CHNO des Quinze-Vingts, INSERM-DGOS CIC 1423, 28 rue de Charenton, F-75012 Paris, France
[3] Paris Eye Imaging Group, Quinze-Vingts National Eye Hospital, INSERM-DGOS, CIC 1423, 28 rue de Charenton, Paris, 75012, France
[4] Institut Langevin, ESPCI Paris, Université PSL, CNRS, 75005 Paris, France,

*authors contributed equally to this work

Corresponding author: kate.grieve@inserm.fr

**Abstract** – Dynamic full-field optical coherence tomography (D-FFOCT) has recently emerged as a label-free imaging tool, capable of resolving cell types and organelles within 3D live samples, whilst monitoring their activity at tens of milliseconds resolution. Here, a D-FFOCT module design is presented which can be coupled to a commercial microscope with a stage top incubator, allowing non-invasive label-free longitudinal imaging over periods of minutes to weeks on the same sample. Long term volumetric imaging on human induced pluripotent stem cell-derived retinal organoids is demonstrated, highlighting tissue and cell organisation as well as cell shape, motility and division. Imaging on retinal explants highlights single 3D cone and rod structures. An optimal workflow for data acquisition, postprocessing and saving is demonstrated, resulting in a time gain factor of 10 compared to prior state of the art. Finally, a method to increase D-FFOCT signal-to-noise ratio is demonstrated, allowing rapid organoid screening.

**Introduction**

The development of microscopy methods dedicated to cell cultures and explants has transformed our understanding of human biology [1-5]. Imaging cells and tissues outside the body removes the constraints of in vivo imaging, enabling higher spatial resolution and the use of exogenous markers [6, 7], thereby permitting imaging of structures ranging from cells and subcellular organelles [8-14] to single-proteins [15, 16]. If the high-resolution imaging of the 3D structural organization of life has brought valuable information [4, 15], the more recent quantification of the temporal dynamics of cells and tissues, made accessible by live cell imaging, gives functional information and a more physiological view of biological behaviours [17-23]. Many processes including cell division rate [21], proliferation, motility, migration, differentiation, or cell death [24], can be used as biomarkers of the physiology of cells and tissues [22-24]. Deregulations of these processes are often associated with diseases such as cancer [25-27], autoimmune disorders [28, 29], neurological disease [30, 31], and chronic inflammation [32]. Live cell imaging of the functional cell response to physical and chemical stimuli can help understanding pathological mechanisms responses to treatments [33] or stimuli [34, 35] and therapeutic effect [33, 36].

However, for successful live cell imaging experiments, cells and tissues must be studied in a context close to their native environment, i.e. as close as possible to in vivo conditions, in order to obtain meaningful data [37-39]. Temperature, $CO_2$, $O_2$ and $N_2$ levels, and tissue culture medium composition are thus critically important [21, 37-39]. Furthermore, 2D cell cultures are not a satisfactory model for this paradigm since the mechanical stiffness of the glass slide supporting the cells exceeds by 6 orders of magnitude the stiffness of most tissues [40, 41], which has strong consequences on cell organization [42], signalling [43], and fate [44]. Additionally, they do not replicate the physiologically important long-range (>100 µm) chemical gradients and long-range 3D interactions, which for instance play a role in organogenesis as well as in the tumor microenvironment [25]. Explants are not completely satisfactory either, due to the difficulty in obtaining them, especially for human explants, as well as the difficulties in maintaining the explants alive once blood circulation has been sectioned, as the explants are then oxygen and nutrients starved leading to cell death after a few hours [45]. During the last decade, the development of differentiation protocols for growing organoids from human stem cells have been a revolution in human biology modelling [4, 5, 46-49]. Organoids are self-organized and self-sustaining 3D cell structures derived from embryonic stem cells or induced pluripotent stem cells (iPSCs), mimicking the 3D cellular organization and composition of the primary tissues, comprising all major cell lineages in proportions similar to those in living tissue. They replicate biologically relevant intercellular phenomena and restores some of the physiological mechanical parameters and long-range chemical and biological interactions [4, 5, 49]. Besides, patient-derived samples have become accessible in large quantities, offering outstanding opportunities for (rare) disease modelling, drug testing and development, and personalized medicine [4, 5, 49]. Despite recent advances in microscopy, high-resolution imaging of organoids is still complex, particularly longitudinal, live, volumetric imaging, and remains an open challenge [50].

For organoid imaging, and more generally for live cell imaging, fluorescence microscopy has largely prevailed over other imaging methods [51, 52]. Owing to the astounding number of fluorescent probes available, able to label specifically most biological entities, the spatio-temporal dynamics of virtually any structure of interest can be studied [53]. However, the use of exogenous fluorophores can often skew native cell functioning and they are intrinsically not suitable for live cell imaging [21, 52, 54-63]. These methods introduce significant artefacts including phototoxicity, increased DNA replication stress and mitotic defects [21, 52], molecular buffering [64], and displaced physicochemical equilibrium associated with preventing the formation of liquid organelles for instance [65]. The use of genetically targeted fluorophores for live imaging is also limited to a few model organisms in which genetic editing tools are available, and they are difficult to transpose for tailored individual studies. More specifically for live cell imaging of organoids, fluorescence-based strategies are possible [39, 66-68] but cumbersome, expensive to implement for patient-derived organoids, and restricted to a live imaging period of a couple of days at most [53]. The integration of fluorophores is also incompatible with cell therapy and organoid grafting therapeutic options. As a result, non-invasive, label-free volumetric optical microscopy appears as a more natural solution for live cell imaging of organoids and other similar 3D tissues [23, 50, 69].

Label free microscopy consists of using the intrinsic optical properties of biological structures—such as scattering, absorption, and phase contrasts, instead of external probes like fluorophores—and monitoring how they change dynamically, in order to characterize samples [70-72]. Recently, the analysis of the temporal dynamics of such endogenous contrasts enabled an important step to be made towards increasing the specificity of label-free microscopies [23, 50, 73].

Among other types of label-free microscopies, full-field optical coherence tomography (FFOCT) appears particularly suited to live cell imaging of organoids [50] thanks to its high 3D spatial resolution, contrast based on backscattering and phase differences [8, 12, 13], high sensitivity, and high imaging speed [8, 12-14, 74]. Furthermore, temporal quantification of FF-OCT signal over a short time scale, named dynamic FF-OCT (D-FFOCT), has emerged as an invaluable metric of metabolic contrast [8, 50, 69, 73] which has since been linked to subcellular organelles activity [14]. As a result, dynamic and static FF-OCT ((D)-FFOCT), have been used in in vitro and ex vivo studies to specifically resolve most cell types in 3D tissues/organisms [8, 9, 14, 74, 75], to identify different cell stage-states such as senescence and mitosis [10, 26, 28], and to detect subcellular compartments and organelles [28]. (D)-FF-OCT is therefore an appealing microscopy design to drive biology research on unaltered samples at high resolution, and perform live cell imaging in thick samples, including 3D cell cultures.

However, whilst D-FFOCT has been successfully applied in retinal cell cultures such as young non-laminated retina organoids [50], retinal explants [69] and retinal pigment epithelium [14, 69], the lack of precise environmental control (temperature, $CO_2$ level, culture medium composition) on the research setups employed have up to now limited live cell imaging to under 3 hours on the same sample [21, 37-39, 45].

In this work, for the first time, we designed a D-FFOCT module coupled to one of the optical ports of a commercial inverted microscope. This design aims to standardize D-FFOCT and make it more accessible to a large community, and facilitate its coupling with a wide variety of other imaging methods available on commercial microscopes. Our microscope was directly installed inside a level 2 biosafety laboratory (L2) so that patient-derived cell cultures can be imaged over long periods of time. More importantly, we took advantage of this integration in a commercial microscope to demonstrate the use of D-FFOCT in physiological environmental conditions thanks to a stage-top incubator. The use of a motorized translation stage as well as the automation capability of a commercial microscope enabled us to perform an automatic continuous live cell imaging of an entire retinal organoid (RO) over more than 2 weeks without observing any sign of stress [14]. The RO was still alive at the end of this period of time. Finally, we demonstrate a new method for imaging retinal organoids (ROs) larger than 1 mm$^2$, older than 250 days, at a stage when photoreceptor cells are fully differentiated with the presence of outer segments. This represents an increasing of the axial range by a factor of 2 and the transverse range by a factor 4 compared to prior state-of-the-art.

## Results

**Volumetric D-FFOCT imaging of retinal organoids—**A volumetric acquisition was carried out on a 28 days old retinal organoid (RO - d28), with the newly developed D-FFOCT module (see Methods), inside a stage top incubator, as displayed in Fig.1, with sufficient coverage to observe the entire organoid (Fig.1a) whilst keeping a subcellular resolution, both in lateral (Fig.1b) and axial directions (Fig.1b-d).

Three dynamic metrics were calculated from a time series of 512 FF-OCT images (see methods), and were displayed in a hue-saturation-brightness space (HSB) [50]. Red hue indicates faster activity than blue; saturation indicates activity randomness; and brightness captures the axial amplitude, or strength of the activity. Using a different acquisition architecture than in prior work [50], acquisition of 512 images at 100 Hz (5.12 s), data transfer (0.79 s), post processing (1.34 s) and saving (0.53 s) are parallelised resulting in an effective total time of acquisition of 5.12 s per D-FFOCT image (see Methods), resulting in a time gain factor of 10 compared to prior state of the art (see Methods) [50]. Thanks to this acquisition speed, three dimensional (3D) volumetric imaging is carried out on the D28 organoid, covering 400x400x120 µm$^3$ in 2 hours and 10 minutes, with a voxel size of 139x139x1000 nm$^3$ (see Methods). 3D volume rendering is achieved, as displayed in Fig1.f-g. We note that the total time of acquisition also included 2-phase static FF-OCT acquisition and stage translation steps, accounting for a slightly longer acquisition time than 5.12 s per field of view.

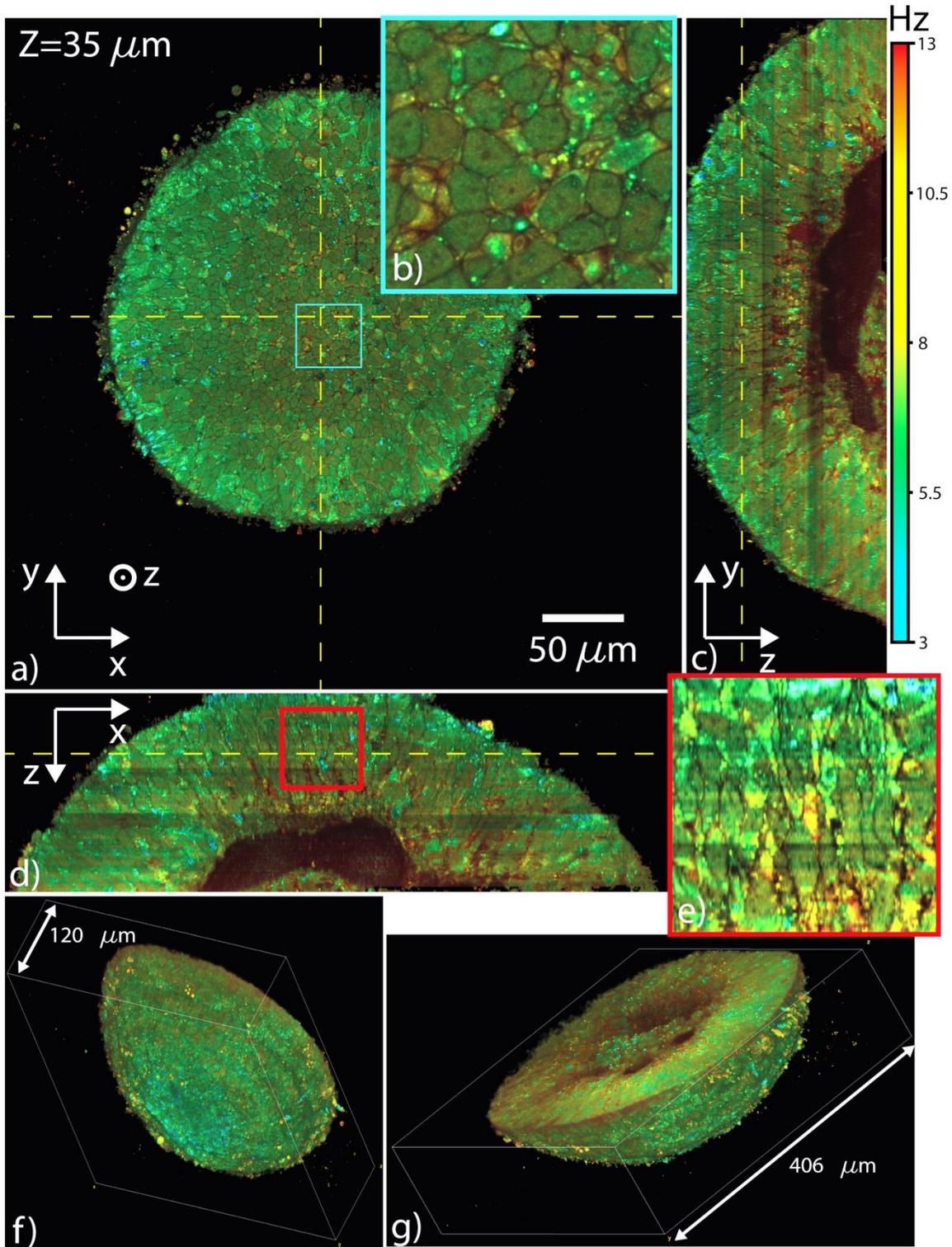

**Figure 1. Example of D-FF-OCT data acquired for each time point of the time-lapse.** A D-FFOCT 3x3 mosaic z-stack of a 28 days old retinal organoid with an axial step of 1 μm is acquired in 150 minutes. Overlap of the tiles is 50%. Hue scales from 3 to 13 Hz mean frequency. Fig.1a shows an *en face* (XY) cross-section from within the organoid, at 35 μm depth. A zoom-in on Fig.1a, indicated by the blue square, is displayed in Fig.1b, where each cell can be resolved. Fig.1c shows a vertical (YZ) cross-section from within the organoid. The vertical yellow dashed lines, displayed in Fig.1a and Fig.1c, coincide. Fig.1d shows a vertical (XZ) cross-section from within the organoid. The horizontal yellow dashed lines, displayed in Fig.1a and Fig.1d, coincide. A zoom-in on Fig.1d, indicated by the red square, is displayed in Fig.1e, where each cells can be resolved along the axial direction. Fig.1f-g show reconstructed volumetric views of the same organoid.

ROs derived from iPS cells are self-forming structures, initially composed of retinal progenitor cells (RPCs) that mimic human retinogenesis through formation of stratified and organized retinal structures that display markers of typical retinal cell types
 [78, 79]. Label free volumetric images confirmed the self-organisation of the structures into a neuroepithelium containing the cell bodies of aligned RPCs (Fig.1c-b), as previously described in studies using classical immunochemistry [79]. At d28, the RPCs appear as confluent green/yellow cells presenting a dynamic profile between 5,5 and 8 Hz in our culture condition.

**Longitudinal and volumetric D-FFOCT imaging of retinal organoids**—A longitudinal volumetric acquisition, with the newly developed D-FFOCT module (see Methods), was carried out on a single retinal organoid, placed inside a stage top incubator, over 17 days. No sign of disturbance or abnormality was observed during the entire acquisition that would indicate RO cellular stress [14, 50], see Fig.2. A D-FFOCT volume, as demonstrated previously and highlighted in Fig.1, was acquired each day with sufficient coverage to observe the entire organoid (Fig.2) whilst keeping a subcellular resolution, both in lateral and axial directions (voxel size of 139x139x1000 nm$^3$). The volumetric acquisition ranged from 2 hours and 10 minutes for d27 to 8 hours 32 minutes for d43, due to the size difference.

In order to make the proof of principle regarding the use of D-FFOCT for time-lapse across several days and weeks, the organoid was embedded in 0.3% Matrigel (see Methods) to keep the same orientation, and "face" imaged throughout the time-lapse. As a result, the position of the organoid was perfectly retrieved after removal and re-insertion of the multiwell plate on the microscope during medium changes. As a consequence, the stage-top incubator on the microscope does not necessarily need to host the sample during the whole period of the time-lapse which can potentially be stored in a separate incubator, freeing up the microscope for carrying out other experiments. The organoid remained in a standard multiwell plate with its unsealed lid during the acquisition, itself placed in a stage top incubator (see Methods) set at 37 C°, 5% $CO_2$, 20.5% $O_2$, 74.5% $N_2$ and 95% humidity. Medium change was carried out every two days under a fume hood placed in a level 2 (L2) biosafety laboratory environment, where the setup itself was installed to insure optimal culture conditions. As a result, contaminations were kept to a minimum, to the standard of L2 and organoid production protocols, throughout the whole experiment.

A selected plane at 50 µm depth originating from the daily volume acquisition, on the same organoid, over 17 days is displayed in Fig.2. Evolution of cells and structures could be followed up over several days at subcellular resolution in the same organoid. Growth of the organoid occurs on a daily basis, thanks to the relatively low concentration of Matrigel embedding the organoid. The shape of the organoid evolves from a spherical ball of retinal progenitors, at d27, to a non-descript shape with rift-like and spherical rosettes, forming from day 36 (d36), which reflects the proliferation of retinal progenitors leading to an increase of neuroepithelial tissue size observed in retinal organoids [78, 79].

During the first week (d27-d35), the RPCs that compose the RO at d27 grow to form the neuroepithelial tissue at the periphery of the RO (Fig.2). Within the two first weeks, rosette formations take place in the neuroepithelium as described previously [78, 79]. We imaged cell motility that lead to the formation of the rosette structures in ROs (Fig. 3). As a result, we could track in 3D their formation and migration over time. We showed that the rosette lumen came from the reorganization of the peripheral layer of the neuroepithelium, implicating hinge regions, as observed in ROs passing from the optic vesicle to the optic cup stage [79]. As a result, proliferative RPCs generated rosette structures which were included in the neuroepithelium over time (Fig.3).

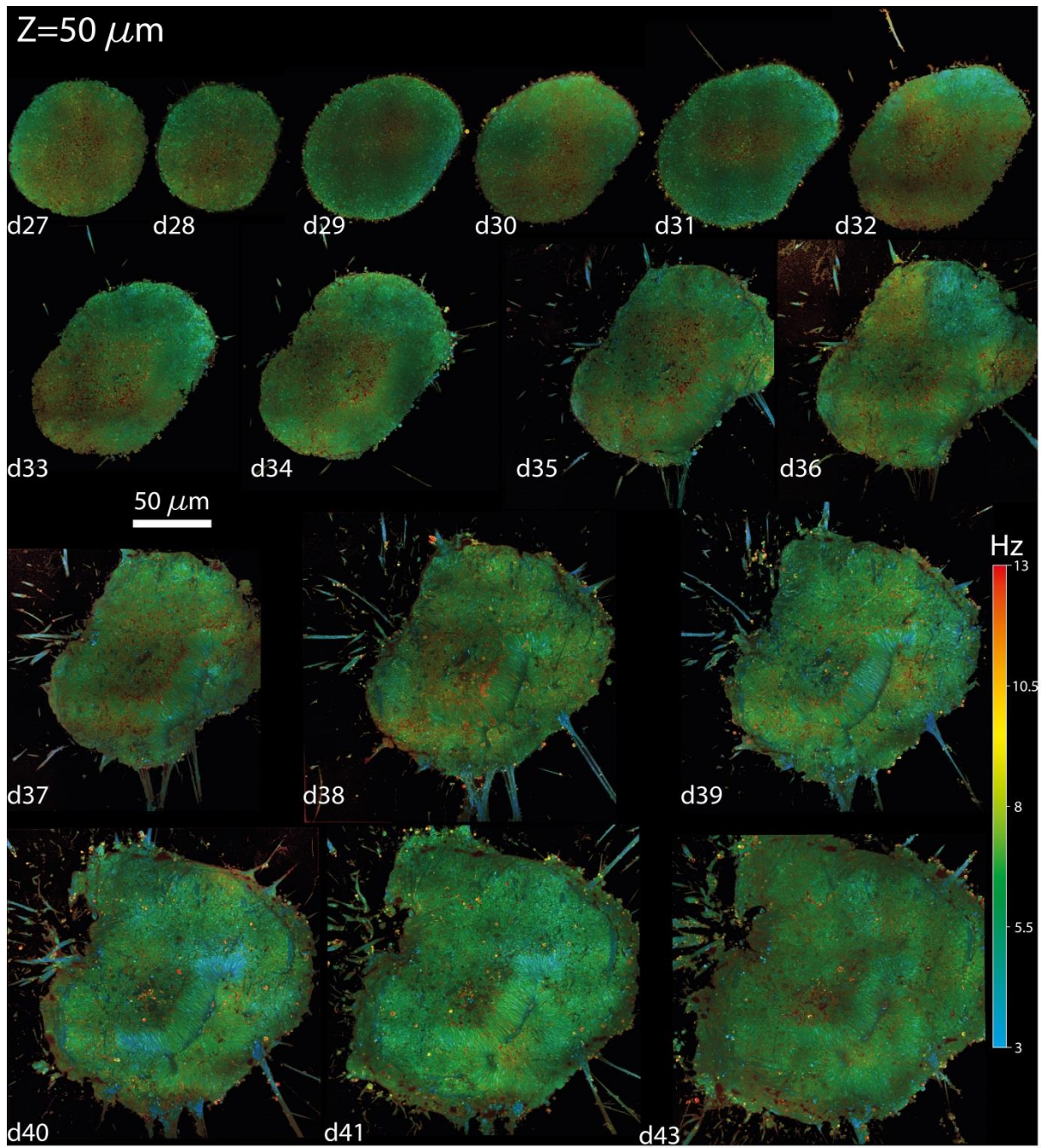

**Figure 2. D-FFOCT volumetric and longitudinal imaging of one single organoid at a depth of 50 µm, across 17 days.** Hue scales from 3 to 13 Hz mean frequency. Image mosaicking covers 406x406 µm², 2928x2928 pixels, (3x3), at day 27 (d27) to 717x717 µm², 5163x5163 pixels, (6x6), at day 43 (d43). The scale bar represents 50 µm and stands for all panels.

The mitotic status of RPCs was highlighted by larger spherical cells than regular RPCs, displaying a red and green hue (5,5 to 13 Hz) close to the lumen, as shown in the d33 ROs (Fig. 3) surrounding the rosette, corresponding to a specific phase of the mitosis. In fact, D-FFOCT resolved cells in replication/division stage, as showed in Fig.4, with a single cell passing successively from the prometaphase, the anaphase and the telophase over a time period of 16 minutes. Also starting from d33 around the rosette, low frequency blue filaments at 3 Hz can be observed. Similar larger filaments can also be observed from d35 onwards in Fig.2. These filaments should theoretically correspond to retinal ganglion cell (RGC) axons, which is coherent with the fact that RGC are known to emerge in ROs from d29 [78]. We could also observe that such filaments appeared earlier at the bottom of the organoid, for example at 4 µm above the glass coverslip, where we could detect them from d29. This earlier detection of the axon-like structures at the bottom of the RO, but not on the sides of equatorial plane may indicate that their formation is favoured by the glass bottom of the multiwall plate, on which the organoid was resting (see

Methods). Low depth prolongation of the axon structures was also favoured because the ROs were placed in Matrigel [47].

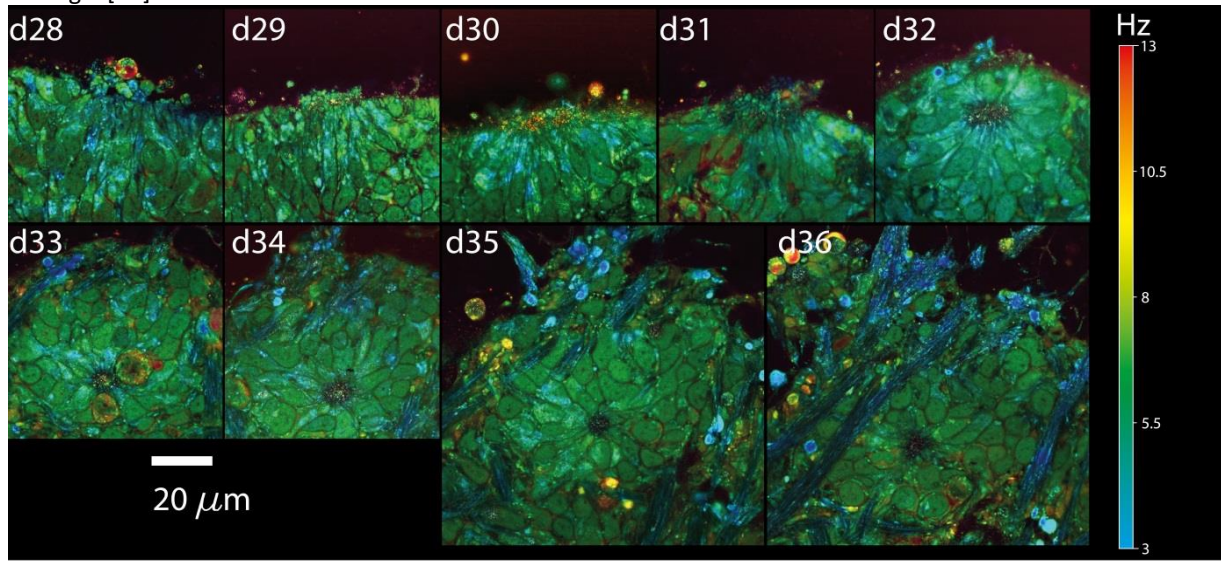

**Figure 3: D-FFOCT longitudinal time tracking of a rosette formation over eight days.** Hue scales from 3 to 13 Hz mean frequency. A single rosette found in the large volume displayed in Fig.2, was tracked throughout each daily volume imaged. All images share the same scaling, with images from day 28 to 34 covering 70x70 µm², 500x500 pixels, and images from day 35 to 36 covering 105x105 µm², 750x750 pixels.

Interestingly, whole images of ROs between d35 to d43 (Fig.2), generated by the D-FFOCT module, showed big rosette structures with long rift-like lumen (Fig.5c), where RPCs are still proliferative, surrounded by the emerging RGCs and retinal inner neurons as described in ROs [78, 79] and shown in Fig.5b. Similar observations were made on different ROs from the same batch on which was carried the longitudinal experiments. Small "punctual" rosettes started to appear from day 32 (d32), mostly from the periphery of the RO, as highlighted by Fig.3, and rift-shaped rosettes started to appear from day 33 (d33), as displayed in Fig.2 and highlighted with Fig.5. The main axis of the cells surrounding rosette was normal to these rifts, as shown by the similarity between Fig.1.e and Fig.5b.

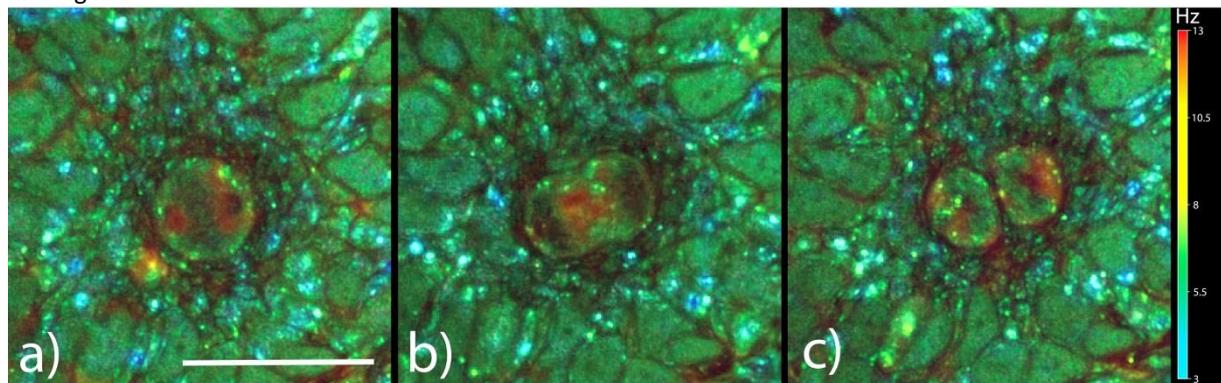

**Figure 4: D-FFOCT tracking of one retinal progenitor cell (RPC) in retinal organoid (ROs) at d36.** Hue scales from 3 to 13 Hz mean frequency. Images cover 40x40 µm², 290x290 pixels. 8 minutes separate each images. Scale bar is 20 µm. Fig.4a highlights a spherical cell in a prometaphase with two spindle poles. Fig.4b highlights an elongated cell characteristic to the anaphase. Fig.4c highlights two daughter cells after the telophase.

Axonal extensions started to be generated inwardly in ROs from d33 (Fig.2). D-FFOCT resolved axonal structures with high resolution showing long associated nerve fibers that can compose the optic nerve in vivo (Fig. 5d, e).

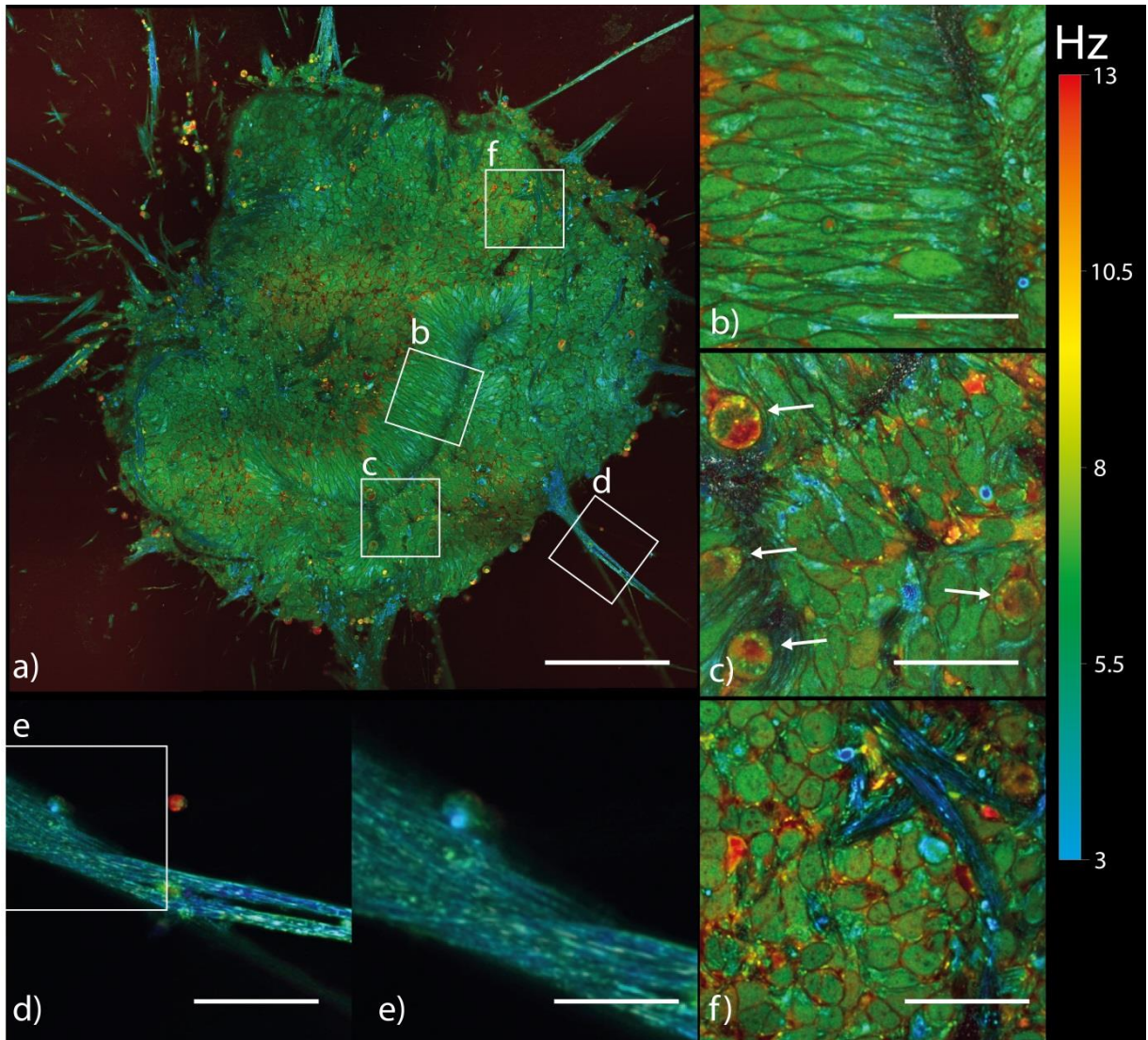

**Figure 5: Highlights on different structures of a retinal organoid (d38) using D-FFOCT.** Hue scales from 3 to 13 Hz mean frequency. Fig.5a displays a single plane mosaic (6x6) of 717x717 µm², 5163x5163 pixels and its scale bar is 140 µm. Fig.5b-d and Fig.5f are magnified views of Fig.5a of 79x79 µm², 570x570 pixels and their scale bar is 30 µm. Fig5.e is a magnified view from Fig.5d covering 35x35 µm², 250x250 pixels, and its scale bar is 11 µm. Fig5.b highlights normal-oriented retinal progenitor cells (RPCs) which are paraxial to the imaging plane. If the RPC can be described analogously to a 3D ellipse, it means that its major axis is included in the imaging plane. Fig5.c highlights mitotic RPCs located nearby the rift-like rosettes. White arrows in Fig.5c point out specifically mitotic RPCs. Fig.5d shows a retinal ganglion cell axon-like structure, expending outwardly from the retinal organoid. Fig.5e is highlighting that our setup can resolve single fibres composing this axon-like structure. Fig.5f is highlighting axon-like structure from retinal ganglion cell propagated inside the organoid.

Temporal evolution of rift-like rosettes could be followed over time too, with an epithelium organisation highlighted with dashed lines on Fig.6. Interestingly, the main cell axis of the RPCs was observed to be radial to the rosette. If RPCs can be described analogously to a 3D ellipse, it means that their major axis is included in the imaging plane. By symmetry, imaging along this axis recapitulates the different aspects of a RPC in a single imaging plane. By extension, since the neuroepithelium is also growing radially to the rosettes, imaging this tissue along their main axis enables to recapitulate both RPC and neuroepithelium fate. Therefore, the alignment of the main cell axis with the D-FFOCT imaging plane enables visualization in a single image of the tissue mesoscopic architecture whilst resolving its cells units. This is a more effective way to monitor RO epithelium as it requires only a single plane rather than a volume to decipher ROs epithelium state.

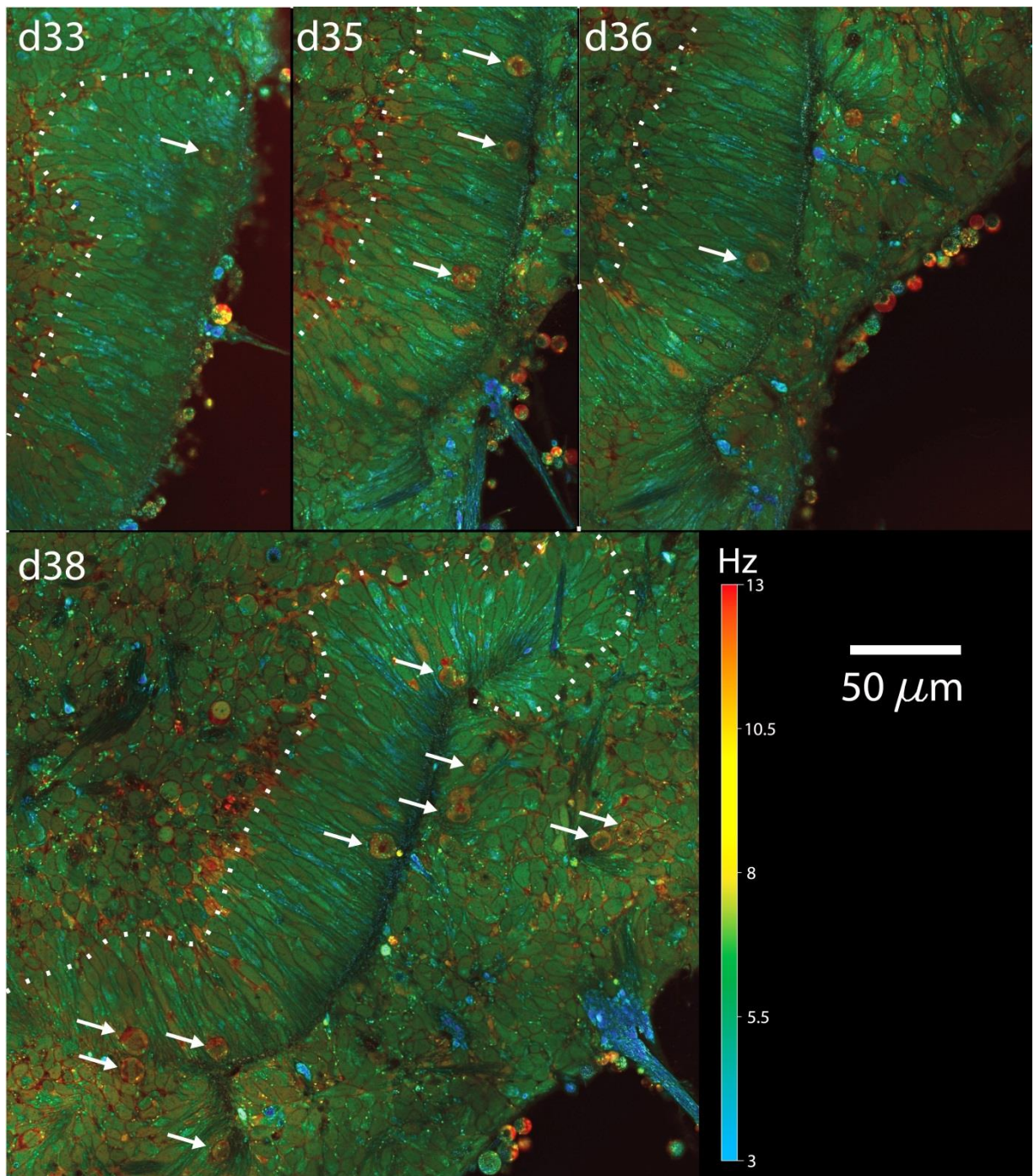

**Figure 6: D-FFOCT longitudinal time tracking of a rosette-rift formation over a couple of days.** Hue scales from 3 to 13 Hz mean frequency. A single rosette-rift was tracked throughout each daily volume imaged. All images share the same scaling with images from day 33 to 34 covering 128x238 µm², 924x1713 pixels, and images from day 36 covering 205x273 um2, 1480x1968 pixels, and day 38 covering 313x296 µm², 2252x2128 pixels. White arrows highlight mitotic RPCs.

In order to demonstrate faster-paced longitudinal acquisition, a time-lapse on a locked plane was carried out over 11 hours, with mosaic imaging (3x3) and larger time series of 1024 FF-OCT images (see Methods), producing a reconstructed D-FFOCT image every 100 seconds (Not shown). This ability shows that faster time monitoring could be conducted on a single plane. However, D-FFOCT single plane imaging at 50 µm depth does not encompasses the general state of the organoid, as mentioned earlier. Therefore, imaging at higher depth within more conventional ROs, with proper layering, would statistically enable the capture of more cells in a radial position. This principle is highlighted in Fig.5b and Fig.6, where radial cells normal to rosettes can be observed for example.

**Improvement of D-FFOCT imaging range for retinal organoids**—D-FFOCT imaging of retinal organoids at high resolution, using objectives with a numerical aperture (NA) higher than 0.8, has been limited to a depth of 80 µm, using a source at 660 nm, in previous work [50]. However, conventional free-floating retinal organoids are spheres that quickly reach a diameters >700 µm during their growth [78, 80]. Since retinal organoids replicate long-range (>100 µm) cell gradients within the retina [25, 78], it is thus critical to be able to image up to a minimal depth of 100 µm in order to capture the overall cell-type organisation in organoids, or mesoscale features, using D-FFOCT [47]. Although we show D-FFOCT images from a RO at 120 µm depths, reaching to the hollow center of the organoid, using a light source at 730 nm (see Fig.1c-d), it requires a significant amount of time to acquire a volumetric dataset, enabling observation of mesoscale features, as displayed in Fig1.e. Furthermore, D-FFOCT images at >100 µm depth arguably lack sufficient SNR to resolve each individual cell and their subcellular features. Ideally, in order to capture both mesoscale tissue cell organisation and subcellular features in a minimum amount of time, imaging would need to be occurring at the "equator" of these spherical organoid, where the radial structures of the organoid align with the *en face* D-FFOCT imaging plane. As a result, higher SNR is needed to optimize the data acquisition on ROs using D-FFOCT, as well as mosaicking over 1 mm$^2$ area on free-floating organoids.

In this section, we demonstrate *en face* imaging of a free-floating organoid over 1 mm$^2$, up to 230 µm depth, on older than 250 days ROs, using a source with a longer wavelength of 810 nm and higher power (20 mW), resulting in an increase of SNR (see Methods). Optimal sampling frequency has been established in the past at 100 Hz for ROs, resulting in a lower D-FFOCT signal at higher or lower sampling frequency [50]. However, here, we used an acquisition rate of 500 Hz, while maintaining the camera full-well-capacity (FWC) near maximum, that we recast in an effective time series at 100 Hz, by performing a temporal binning of 5 successive FFOCT images. As a result, the SNR of the time series was increased by a theoretical factor of $\sqrt{5}$. As a result, longer wavelengths and temporal binning of the data lead to higher SNR level in organoids, enabling imaging at depths up to 230 µm.

Using this acquisition scheme, we were able to image mesoscale tissue features (Fig.7a-b) and sub-cellular details (Fig.7c-f) of much older ROs on a single D-FFOCT mosaic (10x10) image in 12 minutes. In this type of acquisition, different dynamic profiles can be observed at sub-cellular level (Fig.7c-f). As previously reported by fixed organoid approaches (whole mount or cryosections [85,86]), our strategy highlights the layered organization of the organoid cells with different cell types (Fig.7a-b). This approach allows the detection of long filaments of approximately 50 µm in length, located around the edge of the organoid (Fig.7b, 7f), which likely correspond to the inner and outer segments of the photoreceptors. The outermost layer of the retinal organoid, which corresponds to the putative photoreceptor outer nuclear layer, is well defined with very distinguishable cells (Fig.7c). This layer seems delimited towards the internal part of the organoid by a putative outer plexiform layer, mostly present in the whole observation plane (Fig.7a-b). This is followed up with cells presenting different mesoscale features composed of yellow nucleus cells and larger red cells (Fig.7d), which may correspond to the soma of bipolar cells and Müller glial cells that were previously identified by immunostaining in organoids of the same age [85,86]. Finally, in the innermost part of the organoid, distinctive speckled and saturated yellow cells appear (Fig.7e) that corresponds to dying cells, as previously described within the centre of large organoid [86,87] and established with D-FFOCT [14].

Although, not imaged at the "equator" of the organoid, D-FFOCT mesoscale imaging was significantly improved compared to D-FFOCT imaging at lower depths. As a result, ROs can be monitored at a faster rate than with volumetric imaging while encompassing the general state of the retinal organoid.

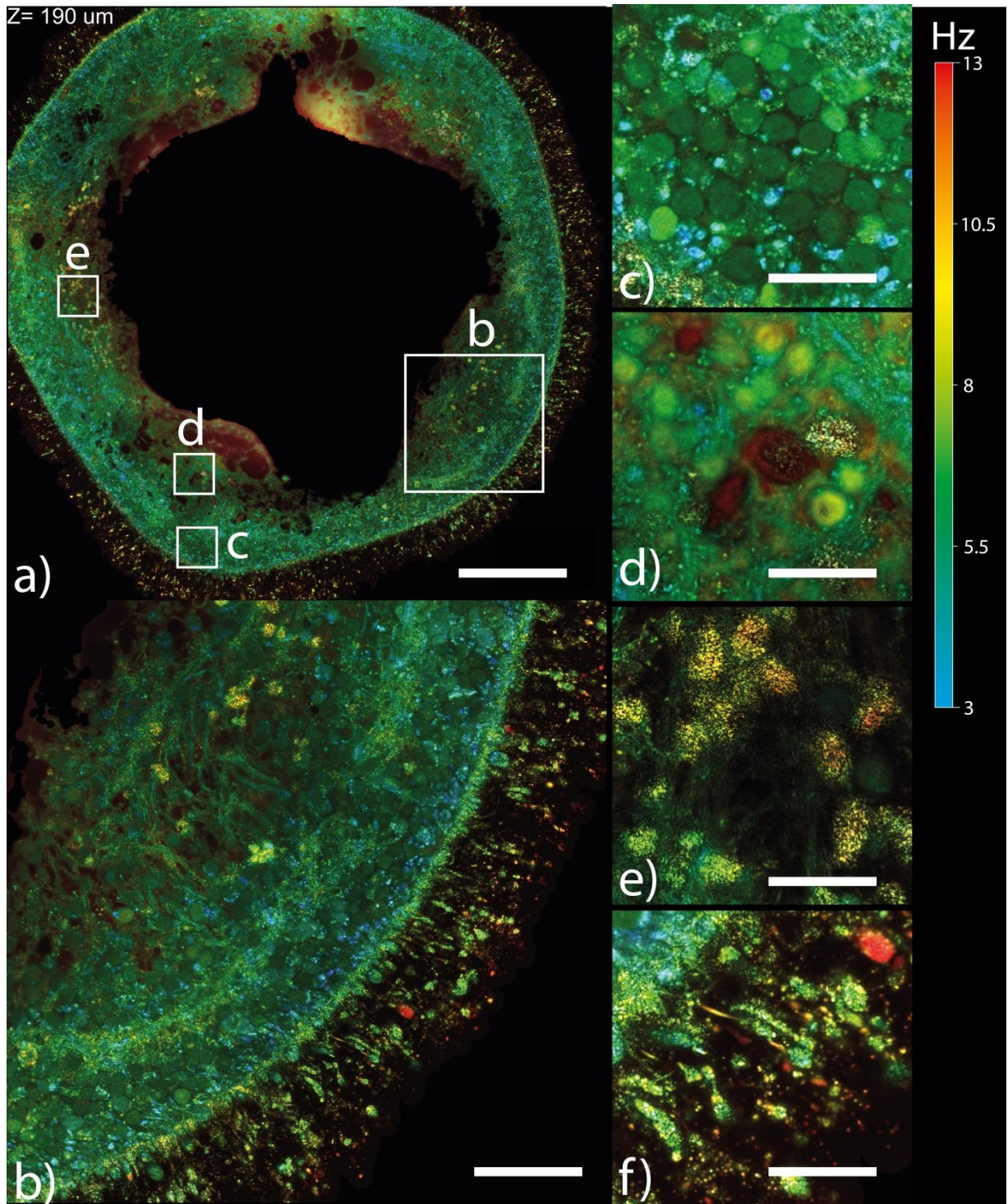

**Figure 7. D-FFOCT large scale mosaic imaging of a retinal organoid highlighting mesoscale and subcellular features.** 266 day old retina organoid being imaged at 190 µm depth by D-FFOCT—10x10 tiles mosaic with 50% overlap, corresponding to 1.125x1.125 mm$^2$, 8101x8101 pixels—is displayed in Fig.3a. Hue scales from 3 to 13 Hz mean frequency. Fig.7b displays a zoom-in of Fig.7a, highlighting the cells layering organisation of the organoid with different cell types. Outer and inner segments of photoreceptors are located around the edge of the organoid (50 µm thick layer), Fig.7a-b, highlighted in Fig.7f. The outer nuclear layer of the photoreceptors is well defined with very distinguishable cells, as highlighted in Fig.7c, and delimited by the outer plexiform layer, mostly present in the whole plane. Fiber-like structures follow up with yellow nucleus cells; see Fig.7d, intertwined within these fibres; which may be bipolar cells. Finally, mostly in the inner part of the organoid, distinctive speckled and saturated cells appear which may be dying or dead cells, see Fig.7e. Scale bar is 60 um in Fig.7b. Scale bar is 25 µm in Fig.7d-f. Scale bar is 10 µm in Fig.7c.

**Imaging in ex vivo tissues—** To evaluate the performance of D-FFOCT to image the highly characteristic physiological organization of photoreceptors, ex vivo adult pig retinal explants were also imaged under culture conditions. Focusing on the photoreceptor layer, images obtained clearly show the cone and rod photoreceptor mosaic in en face slices, with subcellular detail of mitochondria visible inside the cell bodies, and the reconstructed depth slice revealing photoreceptor shape including the characteristic cone inner and outer segments (Fig 8).

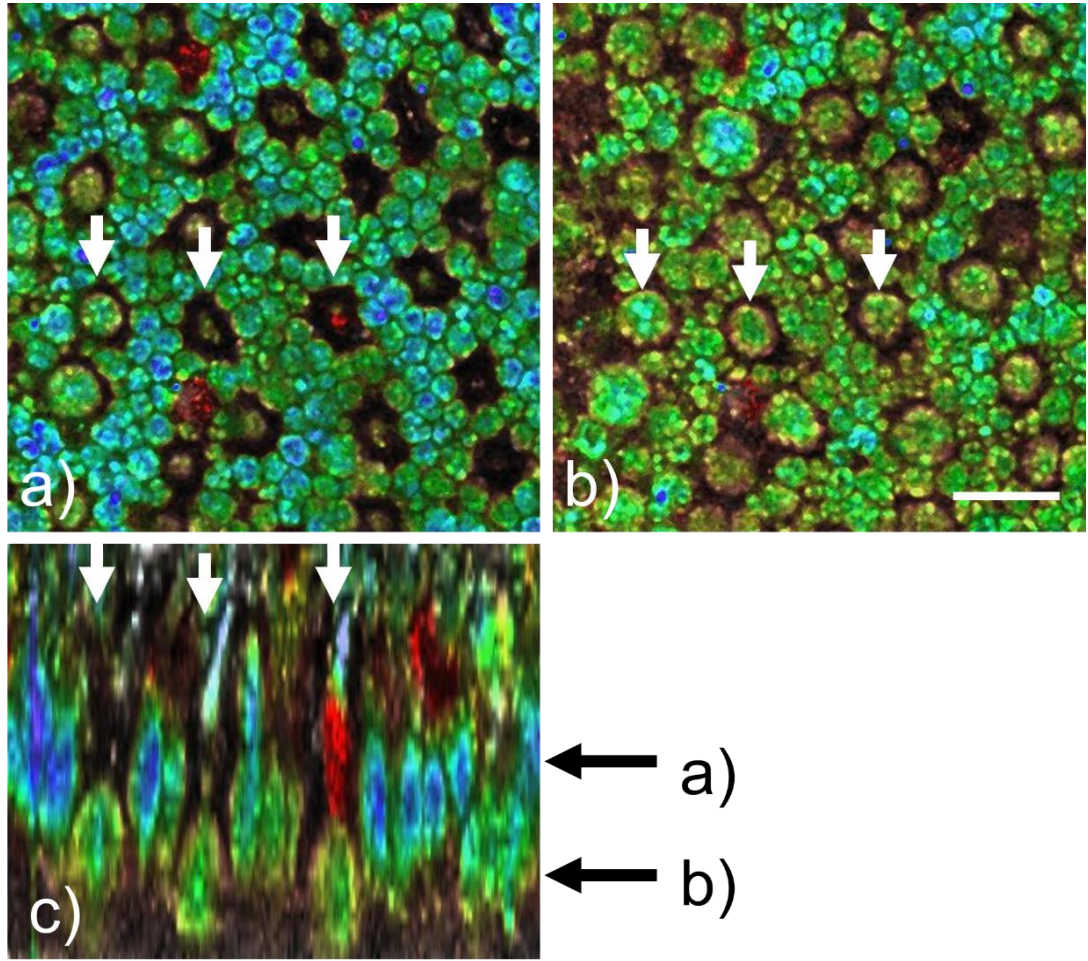

**Figure 8: D-FFOCT in the photoreceptor layer of a porcine retinal explant, imaged under culture conditions.** The cone and rod photoreceptor mosaic were revealed in en face (a, b) and axial (c) slices. The depth positions of a) and b) are indicated by the black arrows next to c). White arrows highlight three cones which can be visualized in all three views. Scale bar, 10µm.

## Discussion

Live imaging with D-FFOCT achieves non-invasive label free 3D viewing of in vitro and ex vivo samples. In this work, we have demonstrated a D-FFOCT module design, coupled to a commercial microscope with stage top incubator, which allowed 3D, longitudinal imaging in retinal organoids over periods of weeks. Use of longer wavelength LED illumination and time series binning allowed acquisition at deeper penetration depths than had been previously demonstrated [50]; and use of the commercial microscope's translation stage allowed coverage of areas larger than 1 mm$^2$ while conserving a theoretical lateral spatial resolution <400 nm [10]. These two capabilities open up D-FFOCT applications on thicker and larger samples, as well as the imaging of retinal organoids in a more equatorial plane. This last aspect is especially interesting as it enables simultaneous monitoring of all the cell layers of the retinal organoid at high resolution, therefore providing an overview of the organoid's overall state in a single D-FFOCT mosaic plane rather than having to make a volumetric acquisition, which implies longer acquisition times and data volume. As a result, control quality and general assessment can be achieved in a quicker way than volumetric acquisition. Live imaging can therefore also be conducted at a faster pace. Furthermore, we note that mosaics were reconstructed from tiles overlapping by a factor of 50%. This inefficient coverage is due to D-FFOCT signal inhomogeneities which occur with the high NA objectives used in D-FFOCT, as sample and reference coherence planes do not overlap perfectly in space. Additional degrees of freedom for controlling the reference mirror inclination (see Methods) independently from the objective inclination would enable correction of this defect. As a result, using an overlap of 10 % instead of 50% for a mosaic becomes feasible. As a result, the area covered by a 10x10 mosaic can be accomplished by a 6x6 mosaic with 10% overlap leading to a threefold time gain. Furthermore, ROs are empty in their center, meaning that this area could be skipped, reducing further the number of tiles necessary to create a large-scale mosaic at this resolution. Additionally, active control of the reference time delay at high depth would benefit the mosaic image quality by having a more homogeneous contrast and higher global SNR. Indeed, the time delay between the reference arm and the sample arm (see Methods) was not constant at a given depth for different x,y coordiantes. The old ROs' heterogeneities cumulated significantly at depths thicker than 100 µm resulting in a significant optical path difference depending on the same plane location. The mismatch was on the order of 1-2 µm on the same plane. Finally, parallel management of data and optimisation of post-processing algorithms enabled close to 10 times faster acquisitions than previously demonstrated [8, 14, 50, 69], critical and paramount for imaging large samples, volumetric imaging and achieving live imaging.

Use of the commercial microscope's translation stage allowed accurate mosaicking over wide regions to cover larger organoids than previously demonstrated, and positioning was reproducible when moving from well to well to the extent that the same field could be recovered without the need for image registration when moving away from and back to the same sample or removing and resetting the plate in position for a subsequent acquisition, when using embedding Matrigel. In the case where the sample is free-floating, slow stepping of 2 µm every 40 ms was carried out in order to minimize movement. No drift was monitored on retinal organoids older than 250 days, see Fig.7. As a result, D-FFOCT can be used for monitoring organoids' production using different protocols, both with or without embedding.

Additional demonstration of this D-FFOCT module was carried out on retinal explants with a focus on photoreceptors, see Fig.8, in order to show imaging performance in ex vivo tissues. To the best of our knowledge, these images represent the highest resolution on photoreceptors achieved by an OCT technique.

Where past work used a standalone D-FFOCT device under optics laboratory conditions to highlight 3D imaging in organoids [50] and cell behavior under stress in the context of disease modelling [14], here, we were able to extend the potential of this promising technique by adapting it to use by a wider community of biologists and other non-optics experts. This was achieved by integrating D-FFOCT to a conventional microscope setup, familiar to biologists, thereby facilitating adoption; allowing continuous protection of the samples under controlled environmental conditions suitable for culture; and by precisely automating key aspects of acquisitions including repeatable xyz positioning, 3D stack capture, and mosaicing; along with developing an efficient and fast workflow for acquiring, post-processing and saving datasets. In addition to these protocol improvements, in terms of optical innovation, we showed greater penetration depths achievable with longer wavelength illumination and the improved SNR gained by binning, which opens up the perspective of making faster acquisitions with fewer frames required to calculate dynamic metrics than in past work. Biological results shown include clear 3D views of cone and rod photoreceptors in retinal explants with unprecedented resolution for an OCT-based technique (see fig. 8) ; two types of iPSC-derived ROs with cell layering distinguishable from the dynamics and morphology of the D-FFOCT signal ; and long term acquisitions on single organoids over >40 day periods, with the ability to probe mature organoids at later stages of development enabled by the wide field mosaicking and deeper penetration depths.

In follow up work, we anticipate the use of the D-FFOCT module in disease modelling and drug screening applications, in 3D samples such as organoids and explants, as well as 2D samples such as cultured cell sheets. A challenge currently being tackled is the efficient management of the large datasets generated by this high resolution volumetric imaging method. Various post processing methods are being explored to extract pertinent data and metrics in an efficient manner, including the use of machine learning based methods to quantify cell density, morphometry and dynamic profile in order to facilitate automation of image analysis [9]. With increased automation of acquisition and image analysis, D-FFOCT may find its place as a tool of choice for live imaging in therapeutic screening and quality control trials of cultured tissues.

## Methods
### Setup
A schematic of the optical setup is displayed in Fig.9, with its components described. The D-FFOCT module is a Linnik interferometer with one additional lens in the sample arm and balanced in the reference arm; see to L3 and L4 in Fig.9. This configuration enables an efficient and homogeneous illumination with a multiport microscope as the distance between the objective in the sample arm (Obj.1) and the microscope output port, where no optics can be placed without altering the microscope hub for other imaging techniques, is relatively important for a D-FFOCT optical layout [8]. This rather unconventional configuration allows to image incoherent reflections, predominantly (99.9%) coming from the reflections on the non-polarizer-beam-splitter (NPBS) cube, in their Fourier plane located at the cMOS camera (Q-2HFW, Adimec, Netherland), reducing their contribution, compared to a classical configuration [74] (see patent PCT/EP2011/066132). The cMOS camera is a Q-2HFW (Adimec, Netherland), with a pixel size of 10x10 µm$^2$ and 1440x1440 pixels, and is set to obtain a SNR level of 1071. Two different LEDs were used in this work, with the central emission wavelengths at 730 µm (LED730) and 810 µm (LED810) (M730L5 and M810L3, Thorlabs, Newport, NJ, USA). The optical elements in the sample, reference and detection arms gives a magnification M= 58 and an imaging field of 200x200 µm$^2$., thus a pixel size of 139x139 nm$^2$. The power applied on the sample is 3.3 and 20 mW, resulting in values of intensity of 53 mW.mm$^{-2}$ and 318 mW.mm$^{-2}$, for the LED730 and LED810, respectively. The exposure time is set at 3.9 ms or 1.3 ms to achieve 95% of the camera FWC, for LED730 and LED810, respectively. In the case of LED730, a time series of 512 images is acquired at 100Hz and is used to generate 3 metrics: the average of the power spectral density (PSD) frequency, the standard deviation of the PSD frequency, the mean of the running standard deviation, with a sliding window of 50 elements— according to a methodology developed by Scholler et al. (2020), for contrast standardisation as well as comparison [50]. In the case of LED810, a time series of 2560 images is acquired at 500Hz and binned in groups of 5 consecutive frames before generating the same metrics as for the case LED730. The three dynamic metrics (mean frequency of the PSD, standard deviation of PSD and averaged running standard deviation) are combined in a HSB space, respectively.

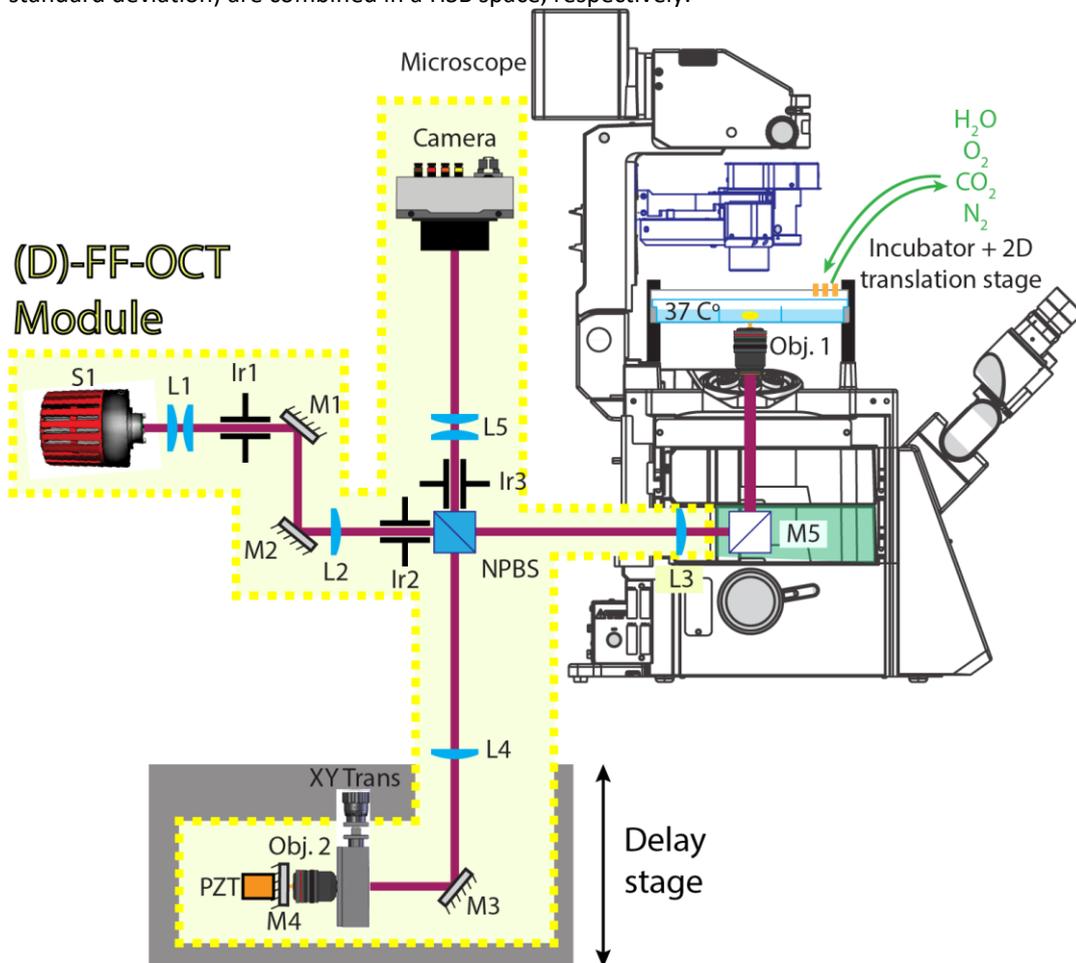

**Figure 9. Schematic of the set-up used in this work**. A mounted light-emitting diode (LED) (M810L3 or M730L5, Thorlabs, Newport, NJ, USA) with a wavelength of 810 nm and 25 nm bandwidth, or 730 nm and 40 nm bandwidth, is used as an extended source (S1) of 1 mm$^2$ with 61.8 µW.mm$^{-2}$, or 13.1 µW.mm$^{-2}$ irradiance, respectively. A first pair of air doublet lenses (L1) is used (AC254-030-B-ML and AC254-150-B-ML, Thorlabs, Newport, NJ, USA) to image S1 onto a first diaphragm (Ir1). A pair of silver mirrors (M1 and M2) (PF10-03-P01, Thorlabs, Newport, NJ, USA) are used to steer the light into the microscope. An air doublet (L2) (AC254-100-B-ML, Thorlabs, Newport, NJ, USA) is used to image Ir1 onto the back focal plane of both objectives (Obj.1 and Obj.2) (UPLSAPO30XSIR, Olympus, Japan) by the intermediate of an air doublet (L3 and L4) (AC254-150-B-ML, Thorlabs, Newport, NJ, USA), respectively. A second iris (Ir2) is imaged onto the sample and M4, a silicon mirror (monocrystalline silicon wafer, Nanomaterials Development Experts Store, China) and is conjugated to a third iris (Ir3), by the intermediary of a non-polarizer-beam-splitter (NPBS), itself conjugated to a cMOS camera (Q-2HFW, Adimec, Netherland) by the intermediary of a pair of air doublet lens es(L5) (AC254-060-B-ML and AC254-200-B-ML, Thorlabs, Newport, NJ, USA). A steering mirror (M3) (PFE10-P01, Thorlabs, Newport, NJ, USA) and a manual translation stage (XY Trans) are used for alignment and fine positioning of Obj.2; standing vertically. A piezoelectric (PZT) (PK25LA2P2, Thorlabs, Newport, NJ, USA) and linear stage (Delay stage) (X-LSQ075A, Zaber, Canada) are used to introduce a fast or fixed phase shift between interferometric fields, respectively. A flat mirror (M5) (CCM1-P01/M, Thorlabs, Newport, NJ, USA) is mounted in the turret (IX3-RFACA-1-3, Rapp Optoelectronic, Germany) of a commercial microscope (IX83, Olympus, Japan). A commercial stage top incubator (H201-K-FRAME, H201-MW-HOLDER and OBJ-COLLAR-2532, Okolab, Italy) enables to control the temperature, humidity, $CO_2$, $O_2$ and $N_2$ gas composition whilst hosting conventional bottom-glassed multiwell plates. A 2 dimensionals translation stage (SCANplus IM 120x80, 2 mm, Marzhauser, Germany) enables scanning from well to well, as well as mosaicking, and the objective holder of the microscope, mounted on a linear stage (U-MCZ-1-2, Olympus, Japan) enables depth (Z) scanning.

**Optimization of data workflow (acquisition, postprocessing and saving)**— Acquisitions are triggered through hardware using a transistor-transistor logic of 0-5 V, generated by an acquisition card (NI cDAQ-9174, National Instruments,TX, USA), and data from the camera (11 bits) are transferred via a 4 CoaxPress cables (25 Gbit/s) on a frame grabber (Cyton CXP4, Bitflow, Massachusetts, USA), itself connected to a PCIe 3.0 (32 Gbit/s) of a PC motherboard (WS X299 SAGE, ASUS, Taiwan). Data from the camera are converted to 16 bits by the frame grabber and are then transferred on a parallel thread of the random-access memory (RAM) (8x Vengeance LPX 1X16GB 3000MHZ, Corsair, California, USA) of the motherboard, and logged onto this same thread. Using a multi-threaded architecture, the effective acquisition time $T_{tot}$ is bottlenecked to the irreducible time needed for data to be logged ($t_{logging}$). For example, for generating 10 D-FFOCT images (H,S,B), from $N_{batches}$ =10 batches, on a locked location, the total time of acquisition is $T^{multithread}_{tot}$=10 x [N x $T_s$ ]+ $t_p$ + $t_{gpu}$ + $t_{save}$  rather than $T_{tot}$=10 x [N x $T_s$ + $t_p$ + $t_{gpu}$ + $t_{save}$ ] as in the previous state of the art [69], where $T_s$ is the inverse of the frame rate of acquisition, N the number of images per batch, $t_p$ the time to transfer the data onto the computing RAM, $t_{gpu}$ the post-processing time on graphical processing unit (GPU) and $t_{save}$ the time it takes to save the post-processed data. The general work flow is sketched in Fig.5. As a result, continuous and lossless acquisitions are possible and the effective generation of dynamic metrics tends ($N_{batches}$→∞) to the incompressible time it takes to acquire the frames themselves—as long as [N x $T_s$] ≥$t_p$, and/or [N x $T_s$] + $t_p$ ≥ $t_{gpu}$, and/or [N x $T_s$] + $t_p$+ $t_{gpu}$ ≥ $t_{save}$—making the acquisition, post-processing and saving limited to $N_{batches}$ x [N x $T_s$]. Whilst non-quantitative D-FFOCT has been demonstrated using Holovibe (http://holovibes.com/) [14, 81], an acquisition program written in native C++ and CUDA, for fast acquisition and data processing, resulting in calculation of a dynamic metric every 160 ms, we note that this method used forecasting methods which works only an a fixed location in the sample and is not compatible with mosaicking nor ZStacking. As a result, our acquisition framework is optimal.

This data framework was accomplished with MATLAB, using the toolbox Imaqtool. A function is triggered every N frames, transferring these N frames using getdata on a second RAM CPU thread, whilst the first thread carries on the logging of another batch simultaneously. The data are then transferred to a GPU (GeForce RTX 3090, NVIDIA) where the *dynamic* metrics are calculated in $t_{gpu}$ = 1.35 s for N=512, approximatively 7 times faster than the previous state of the art [69]. The post-processing of the data was re-written to achieve a speed gain of 3.3. Both scripts were compared on a same GPU with 8 GB of memory for fairer script comparison (GeForce RTX 2070 with Max-Q Design, NVIDIA), resulting in 3.45 versus 12.35 seconds. Saving of the *dynamic* metrics on a HDD (4TO Blue S-ATA III 64Mo, WD) is parallelised on another RAM CPU thread using the function parfeval. Overall, the acquisition, data processing and saving of the dynamic metrics now takes 5.12 s compared to 50.5 s in the previous state of the art [50]. A slower acquisition time of 7.42 s was achieved when mosaicking on free floating organoids, due to the slow translations steps of 2 μm per 20 ms, as the whole data flow had to be stopped and reinitialized during the slow drift of the 2D translation stage.

The microscope functionalities as well as the 2D XY translation stage are interfaced through JAVA functions developed by μManager [82]—in which MATLAB is coded—and the other instruments are controlled through MATLAB. A dedicated graphic user interface (GUI) was created from scratch for this setup.

Mosaics were reconstructed from tiles with 50% overlapping using MIST (Microscopy Image Stitching Tool) [83], a plug-in available on Fiji [84]. For the free-floating organoids, stage acceleration was kept minimal by micro stepping with increment of 2 microns in order to avoid sample drifting.

## Sample preparation
### Human iPSC maintenance and retinal differentiation
Human iPSC line-5f derived from retinal Müller glial cells, was cultured on truncated recombinant human vitronectin-coated dishes and using the mTeSRTM1 medium (StemCellTM Technologies) as previously described [80, 85]. For retinal differentiation, adherent iPSCs were expanded to 70–80%, and FGF-free (fibroblast growth factor) medium was added to the cultures for 2 days, followed by a neural induction period allowing retinal structures to appear. Identified retinal organoids were manually isolated around d28 and cultured as floating structures in ultra-low attachment 24-well plates (Corning) as floating structures in the ProB27 medium supplemented with 10 ng/ml of animal-free recombinant human FGF2 (Peprotech, , 100-18B) and half of the medium was changed every 2-3 days [80, 85]. The ProB27 medium is composed of chemically defined DMEM:Nutrient Mixture F-12 (DMEM/F12, 1:1, L-Glutamine), 1% MEM non-essential amino acids (Thermo Fisher Scientific), 2% B27 supplement (Thermo Fisher Scientific), 10 units/ml penicillin and 10 µg/ml streptomycin. At d35, retinal organoids were cultured in the absence of FGF2 in the ProB27 medium with 10% FBS (Thermo Fisher Scientific) and 2 mM of Glutamax (Thermo Fisher Scientific) for the next several weeks. Around d84, the retinal organoids were cultured in the ProB27 medium with 2% B27 supplement without vitamin A (Thermo Fisher Scientific) until d250 [85, 86].

**RO sample preparation for D-FFOCT imaging**

Early ROs at d28 were embedded in 3% Matrigel (Corning® Matrigel® Basement Membrane Matrix Growth Factor Reduced Phenol Red Free [Corning, 356231]) in 12-well glass bottom plate (IBL, 220.210.042). Embedded structures were cultured in ProB27 medium +FGF2 up to D35 followed by 1 week in ProB27 medium. Half of the medium was changed every 2-3 days [80, 85]. Old ROs around D250 were placed to a black, flat-bottomed glass microscopy-compatible 24-well plate (ibidi) and the well is filled with pre-warmed fresh culture medium. The organoid is left for at least 1h in the dark in the 37°C / 5% CO2 incubator before imaging.

**Retina explant**

Porcine eyes were obtained from a local slaughterhouse in agreement with the local regulatory department and the veterinarians from the French Ministry of Agriculture (agreement FR75105131). Eyes were dissected to isolate retinas in CO2 independent medium (18045054, Thermo), and pieces from the region behind the optic nerve were cut using sterile biopsy punches of 2mm. Each retinal explant was then placed on a polycarbonate membrane (140652, Thermo, Waltham, Massachusetts) photoreceptors turned upwards. This assembly of membrane plus explant was then placed face down on a glass plate (Cellvis, P12-1.5H-N, IBL) for microscopy, i.e. with the explant sandwiched between membrane and plate, and imaged with ganglion cells uppermost. The level of medium was precisely controlled to hydrate the retina without complete immerging it to have optimized oxygenation. These pieces were kept in culture in a CO2 incubator at 37°C for 3 days in Neurobasal-A medium (10888022, Thermo) containing 2mM of L-glutamine (G3126, MERCK).

**Data and code availability**

Due to the large size of the datasets (>1 To), they cannot be uploaded on a public repository. However, data, including the full volumetric longitudinal dataset, the fast timelapse over 11h, and the dataset obtained on the d266 organoid, as well as the codes can be made available upon reasonable request from the authors.

**Code availability**

Data and codes are available upon reasonable request from the authors.

**Acknowledgements**

We thank Anis Aggoun for suggesting using the MIST plugin from ImageJ.

**Author Contributions**

The overall project was conceived and supervised by KG.

Optical design and construction was conceived and carried out by TM supervised by OT. TM conceived and carried out the interfacing, as well as the design and implementation of software architecture and post-processing optimizations (acquisition GUI, processing).

The method for increasing imaging depth and mosaicking with D-FFOCT was conceived and carried out by TM. Acquisition protocols were designed by SR, VF and TM, and acquisitions were carried out by TM, with assistance of SA, on samples provided by SR, OG, AS, SP, and VP, with assistance from JB and MC.

Images and volumes were reconstructed by TM.

KG, TM, OT, SR, OG and SA discussed the results and wrote the article.

**Competing Interests statement**

The authors declare that they have no known competing financial interests or personal relationships that could have appeared to influence the work reported in this paper.